\documentclass[aps,prl,twocolumn,groupedaddress,superscriptaddress,amsfonts,
citeautoscript,a4paper]{revtex4-1}
\usepackage{latexsym}
\usepackage{amsmath}
\usepackage{amssymb}
\usepackage{graphicx}
\usepackage[T1]{fontenc}
\usepackage[open]{bookmark}
\usepackage{hyperref}
\hypersetup{colorlinks=true,allcolors=blue}
\usepackage{orcidlink}
\usepackage{graphicx}
\usepackage{dcolumn}
\usepackage{bm}
\usepackage{xcolor}
\usepackage{hyperref}
\usepackage{cleveref}
\usepackage{amsmath}
\usepackage{float}
\usepackage{svg}
\usepackage{amssymb}
\usepackage{scalerel}
\usepackage{caption}
\usepackage{wasysym}

\makeatletter
\pretocmd\frontmatter@keys@format{\addvspace{20\p@}}{}{}
\makeatother

\usepackage{svg}
\usepackage{cleveref}
\usepackage{siunitx}
\usepackage{caption}

\begin{document}

\title{Sub-to-super-Poissonian photon statistics in cathodoluminescence of color center ensembles in isolated diamond crystals}

\author{Saskia Fiedler\,\orcidlink{0000-0002-7753-0814}}
\thanks{S.~M. and S.~F. contributed equally to this work.}
\noaffiliation

\author{Sergii Morozov\,\orcidlink{0000-0002-5415-326X}}
\thanks{S.~M. and S.~F. contributed equally to this work.}
\noaffiliation

\author{Danylo Komisar\,\orcidlink{0000-0001-8856-7586}}
\noaffiliation

\author{Evgeny A. Ekimov\,\orcidlink{0000-0001-7644-0078}}
\noaffiliation

\author{Liudmila~F.~Kulikova\,\orcidlink{0000-0002-9070-0590}}
\noaffiliation

\author{Valery~A.~Davydov\,\orcidlink{0000-0002-8702-0340}}
\noaffiliation

\author{Viatcheslav N. Agafonov\,\orcidlink{0000-0001-5770-1252}}

\author{Shailesh~Kumar\,\orcidlink{0000-0001-5795-0910}}
\noaffiliation

\author{Christian Wolff\,\orcidlink{0000-0002-5759-6779}}
\noaffiliation

\author{Sergey I. Bozhevolnyi\,\orcidlink{0000-0002-0393-4859}}
\noaffiliation

\author{N. Asger Mortensen\,\orcidlink{0000-0001-7936-6264}}
\thanks{Corresponding author}
\noaffiliation

\begin{abstract}
Impurity-vacancy centers in diamond offer a new class of robust photon sources with versatile quantum properties. While individual color centers commonly act as single-photon sources, their ensembles have been theoretically predicted to have tunable photon-emission statistics. Importantly, the particular type of excitation affects the emission properties of a color center ensemble within a diamond crystal. While optical excitation favors non-synchronized excitation of color centers within an ensemble, electron-beam excitation can synchronize the emitters and thereby provides a control of the second-order correlation function $g_2(0)$. In this letter, we demonstrate experimentally that the photon stream from an ensemble of color centers can exhibit $g_2(0)$ both above and below unity. Such a photon source based on an ensemble of few color centers in a diamond crystal provides a highly tunable platform for informational technologies operating at room temperature.
\end{abstract}

\maketitle

\section{Introduction}

The ever-decreasing sizes of photonic nanostructures and optoelectronic devices have increased the demand of suitable characterization methods at the true nanoscale~\cite{Dombi_RMP_2020,Aharonovich:2016}.
Due to the high spectral and spatial resolution down to the (sub-) nanometer range, electron beam-based techniques, namely cathodoluminescence (CL) and electron energy loss spectroscopy (EELS), have become powerful characterization tools~\cite{GarciadeAbajo:2010,Polman:2019} and extreme near-field probes of complex electrodynamic response~\cite{Nelayah_NatPhys_2007,Duan:2012,Scholl:2012,Raza:2014,Losquin:2015,Campos:2019}.
CL microscopy is widely used for the correlation of structural and optical properties of nano-sized structures~\cite{,Sannomiya:2020,Fiedler:2022,Varkentina:2022}, allowing for nano-mapping of spectral properties and mode characterization, while also providing information about angular emission properties~\cite{Fiedler2020,Mignuzzi2018}. 
CL in combination with the Hanbury Brown and Twiss (HBT) interferometry has also been utilized for investigation of light emission statistics in single-photon emitters~\cite{Tizei:2013,Bourrellier:2016,Feldman:2018} as well as of ensembles of solid-state quantum emitters~\cite{Meuret:2015,Meuret:2017, FiedlerWS2}.

The second-order intensity-correlation function $g_2(\tau)$ is allowing for identification of individual emitters, which exhibit photon antibunching, with a characteristic fingerprint $g_2(0)<0.5$ for zero delay time $\tau$.  
An ideal single-photon emitter is characterized by sub-Poissonian photon statistics $g_2(0)\rightarrow0$, while an ensemble of $N$ emitters exhibits $g_2(0)\rightarrow1-1/N$, thus inevitably approaching Poissonian statistics in the limit of large $N$.
Remarkably, the electron-beam excitation can result in synchronization of emitters in an ensemble, which manifests in $g_2(\tau)$ as photon bunching, i.e. $g_2(0)>1$~\cite{FiedlerWS2}. 
The degree of synchronization of an ensemble of $N$ emitters can be controlled by electron-beam parameters, while the bunching peak provides information on emission lifetime and the probability of the electron excitation to interact with an emitter~\cite{Meuret:2020}. 

\begin{figure}[b!]	
\includegraphics[width=1\columnwidth]{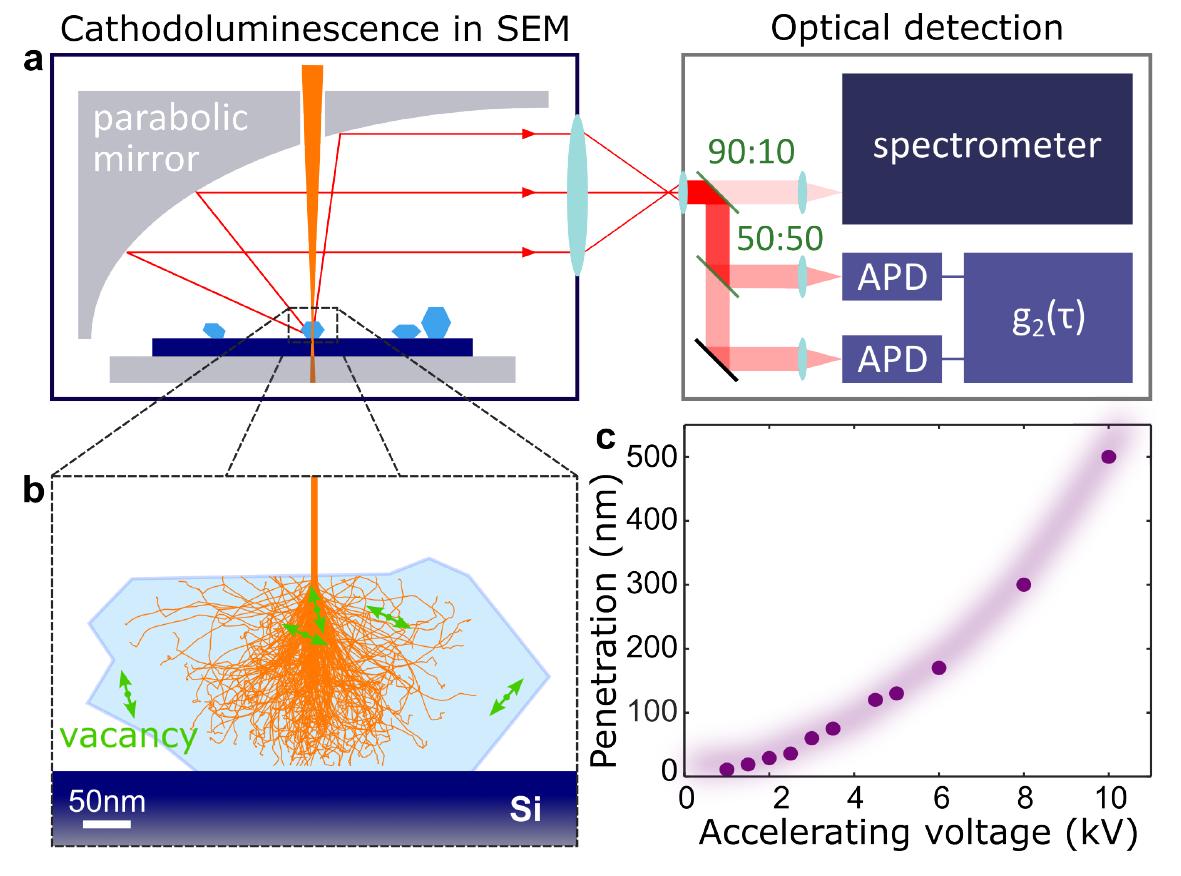}
\caption{\textbf{Electron beam-based correlation spectroscopy of color centers in a diamond. }
	\textbf{a}~Focused electron beam (orange) excites cathodoluminescence of color centers in an isolated diamond crystal (blue) in an SEM chamber. The generated photons (red) are collected by a parabolic mirror and sent to a spectrometer and a TCSPC system. 
	\textbf{b}~Monte Carlo simulation of primary electron trajectories in diamond for a \SI{5}{kV} incoming electron beam with a spot size of \SI{5}{nm}. 
	\textbf{c}~Penetration depth of electrons in diamond (energy loss of 75\%) at increasing accelerating voltages.
	}
	\label{fig-intro}
\end{figure}

To achieve high synchronization of emitters, the excitation of all emitters in an ensemble should be done via the same electron, which can be simply realized at very low electron-beam currents (few sub-pA)~\cite{FiedlerWS2}. 
At such low currents the average time interval between incoming electrons is much longer than the radiative lifetime of emitters in an ensemble. 
In the limit of high electron-beam current, the primary electrons arrive too closely spaced in time to allow for any synchronization, typically leading to a flat second-order auto-correlation function, $g_2(\tau)\to1$, as in the more common case of photo-excitation. 
A single-photon emitter exhibits current-independent photon statistics, $g_2(0)\rightarrow0$, while a large ensemble has current-dependent behavior approaching $g_2(0)\rightarrow1$ at high currents.
In contrast, the intermediate case of very few quantum emitters in an ensemble has been theoretically predicted to exhibit a transition from sub to super-Poissonian statistics via tuning of the electron-beam current~\cite{Meuret:2015,Temnov:2009,Yuge:2022}.
However, the experimental observation of the transition in photon statistics has been hindered by degradation of ensembles at high electron-beam currents in the  investigated so far solid-state platforms~\cite{Meuret:2015}.

In this letter, we report the photon-correlation statistics of germanium vacancy (GeV$^-$) and silicon vacancy (SiV$^-$) ensembles in isolated diamond crystals excited by an electron beam. 
We study the CL response in the limit of a single color center, or a pair of such centers, and eventually the transition to ensembles containing over $10^2$ centers. 
In the case of a single color center excitation, we observe antibunching behavior with 
{$g_2(0) = 0.06$} regardless of the applied electron-beam current; for ensembles we find photon bunching of up to {$g_2(0) = 4.9$} for the lowest attainable probe currents by our CL setup. 
Remarkably, for the case of a few single-photon emitters, we experimentally demonstrate the tunability of the photon emission statistics from sub to super-Poissonian dynamics via ensemble synchronization, i.e., a crossover from $g_2(0)<1$ to $g_2(0)>1$. 

\section{Results and discussion}

\subsection{Electron-beam excitation of color centers}

We investigate  germanium- and silicon-doped diamond crystals synthesized by high-pressure methods~(see SI)~\cite{Ekimov_DRM_2019,Kondrina:2018,Davydov:2014}.
The experimental setup for the CL measurements is schematically shown in Fig.~\ref{fig-intro}a.
Here, the high-energy electrons are focused on a diamond crystal in a scanning-electron microscope (SEM) chamber through a small hole in a parabolic mirror.
The generated CL emission is subsequently collected via the parabolic mirror and sent for characterization with a spectrometer and an HBT interferometer with two avalanche photodiodes (APD), which constitutes a time-correlated single-photon counting (TCSPC) system (see details in SI).

Fig.~\ref{fig-intro}b schematically shows a simulation of a diamond crystal with color centers (green double-sided arrows), which is penetrated by a \SI{5}{\kilo\volt} electron beam. 
The orange lines represent trajectories of primary electrons modeled by Monte Carlo simulation, which are spreading in a \SI{500}{nm}-thick diamond crystal on a Si substrate~\cite{Drouin:2007}. 
We repeat the simulation for a range of acceleration voltages to find an electron penetration depth in the diamond crystal by estimating the electron-energy loss at 75\% (Fig.~\ref{fig-intro}c).
Depending on the size of each studied diamond crystal, the acceleration voltage of the electron beam can be used to control its penetration depth and interaction volume.
In this way, one can systematically locate and isolate ensembles of color centers close to the surface as well as those deeply buried in the crystal ({SI Fig.~S1}). 
The scattered electrons cause high-energy collective interband excitations in diamond (also referred to as bulk plasmons in the electron spectroscopy literature~\cite{Egerton2011,Meuret:2015}), which subsequently decay into multiple electron-hole pairs, which can further radiatively recombine through color centers.

\begin{figure*}	
\includegraphics[width=1.99\columnwidth]{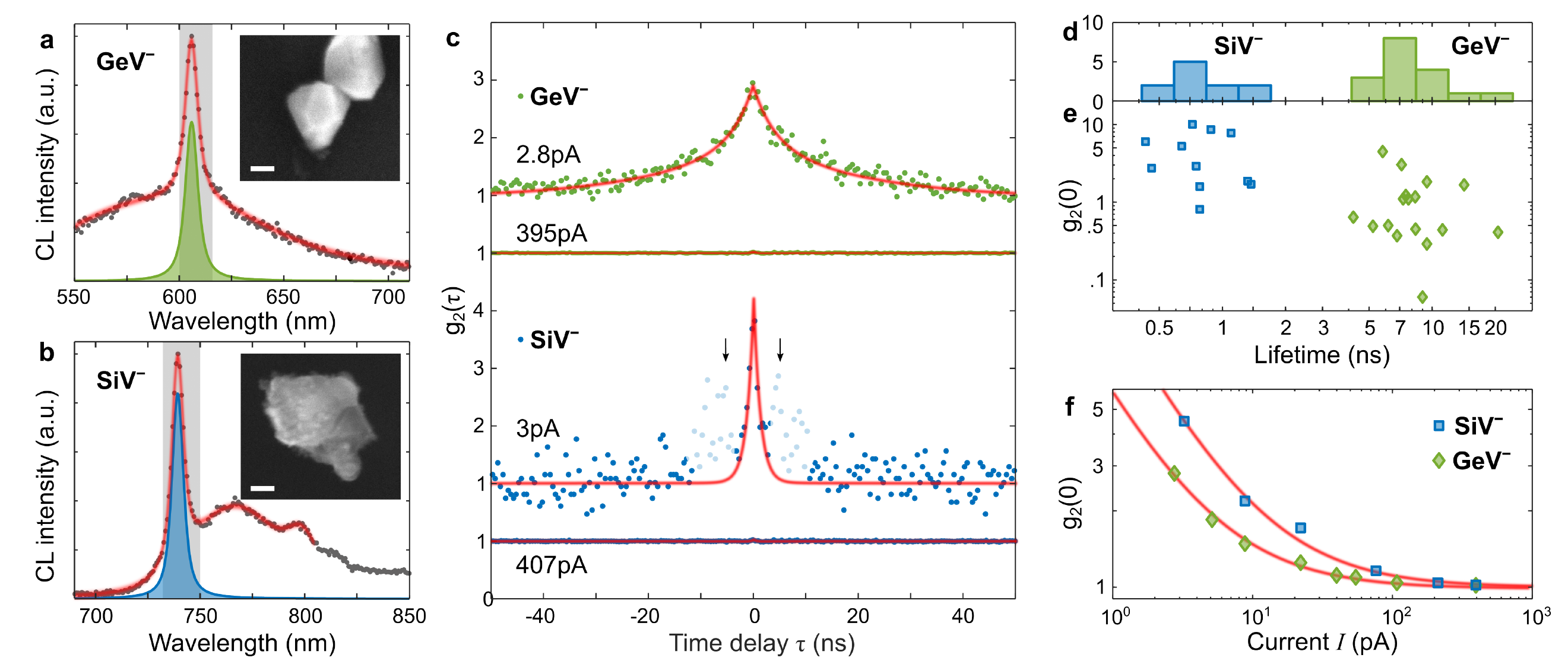}	
\caption{
\textbf{Photon bunching in CL of GeV$^-$ and SiV$^-$ ensembles with large number of color centers ($N>100$).}
    \textbf{a-b}~CL spectra of diamond crystals with ensembles of GeV$^-$ and SiV$^-$ centers. The fit (red) reveals the ZPL at \SI{606}{nm} (green) for the GeV$^-$ ensemble and \SI{739}{nm} (blue) for the SiV$^-$ ensemble over the background diamond emission. The gray-shaded spectral range indicates the transmission of a band-pass filter used to collect intensity-correlation histograms. The insets show SEM images of the investigated crystals. The scale bars denote \SI{100}{nm}.
    \textbf{c}~Photon statistics of GeV$^-$ (green) and SiV$^-$ (blue) emitter ensembles measured in CL at low and high electron-beam currents. The shaded satellite peaks in the SiV$^-$ histogram (indicated by arrows) are afterglow  artifacts of detectors (see SFig.~X). 
    \textbf{d-e}~Emission statistics over 17 crystals with GeV$^-$ centers and 11 crystals with SiV$^-$ centers.
    \textbf{f}~$g_2(0)$ is inversely proportional to the electron-beam current $I$ (red lines) and converges to $1$ at high currents.    
}	
\label{fig-many}
\end{figure*}

\subsection{Large ensembles of GeV$^-$ and SiV$^-$ centers}

The insets of Fig.~\ref{fig-many}a-b present SEM images of representative diamond crystals containing multiple GeV$^-$ and SiV$^-$ centers. To confirm the presence of color centers, we recorded photoluminescence (PL) spectra with characteristic positions of the zero-phonon line (ZPL) for GeV$^-$ and SiV$^-$ color centers (Fig.~\ref{fig-many}a-b).
Spectral decomposition fitting resulted in the ZPL position and linewidth for GeV$^-$ (\SI{606}{nm}, \SI{7}{nm}) and SiV$^-$ (\SI{739}{nm}, \SI{8}{nm}) color centers, which agree well with ZPL parameters typically observed in PL measurements~\cite{Bradac:2019,Takashima2021,Lagomarsino2021,Nahra2021}.
We note that the redshifted emission in the recorded spectra corresponds to the phonon side bands relaxation channel~\cite{Prawer2014}, while the broad background is associated with emission from impurities in diamond as well as Mie-related excitations in diamond crystals~\cite{Fiedler:2022} (see {SI Fig.~S2} for a reference spectrum of a crystal without color centers). 

We measure photon statistics in the generated CL signal with an HBT interferometer. 
We use band-pass filters to isolate the ZPL and suppress background emission (see gray-shaded areas in Fig.~\ref{fig-many}a-b for transmission window). 
Fig.~\ref{fig-many}c presents the second-order intensity auto-correlation function $g_2(\tau)$ of GeV$^-$ (green) and SiV$^-$ (blue) ZPLs acquired at low and high electron-beam currents $I$.
At the low current of few pA, we observe photon bunching, which can be quantified by fitting the $g_2(\tau)$ histogram, extracting $g_2(0)=2.8\pm0.1$ and $g_2(0)=4.5\pm0.1$ for GeV$^-$ and SiV$^-$ ensembles, respectively ({see fitting details in SI}). 
Fits of bunching peaks in Fig.~\ref{fig-many}c also qualitatively confirm the presence of a large number of emitters $N>100$ in both GeV$^-$ and SiV$^-$ ensembles, however the signal-to-noise ratio at low electron-beam currents complicates a more accurate quantitative estimation of $N$. 

The width of the bunching peak is linked to the effective radiative lifetime of the excited state~\cite{FiedlerWS2}, and can be extracted from the $g_2(\tau)$ fit. 
For the ensemble of GeV$^-$ centers, we find a bi-exponential decay behavior with fast and short radiative components of {$7.6$~ns} and {$23$~ns}, while for SiV$^-$ -- there is a monoexponential decay with an effective lifetime of {$1.1$~ns}. 
The long emission component in germanium-doped crystals indicates the presence of a shelving state in the recombination pathway~\cite{Nahra2021}.
We repeated the lifetime measurements acquiring emission statistics (Fig.~\ref{fig-many}d-e), extracting average effective radiative lifetimes of {$7.1$~ns} and {$0.9$~ns} for GeV$^-$ and SiV$^-$ ZPLs under electron-beam excitation, respectively.

\begin{figure*}
\includegraphics[width=0.99\linewidth]{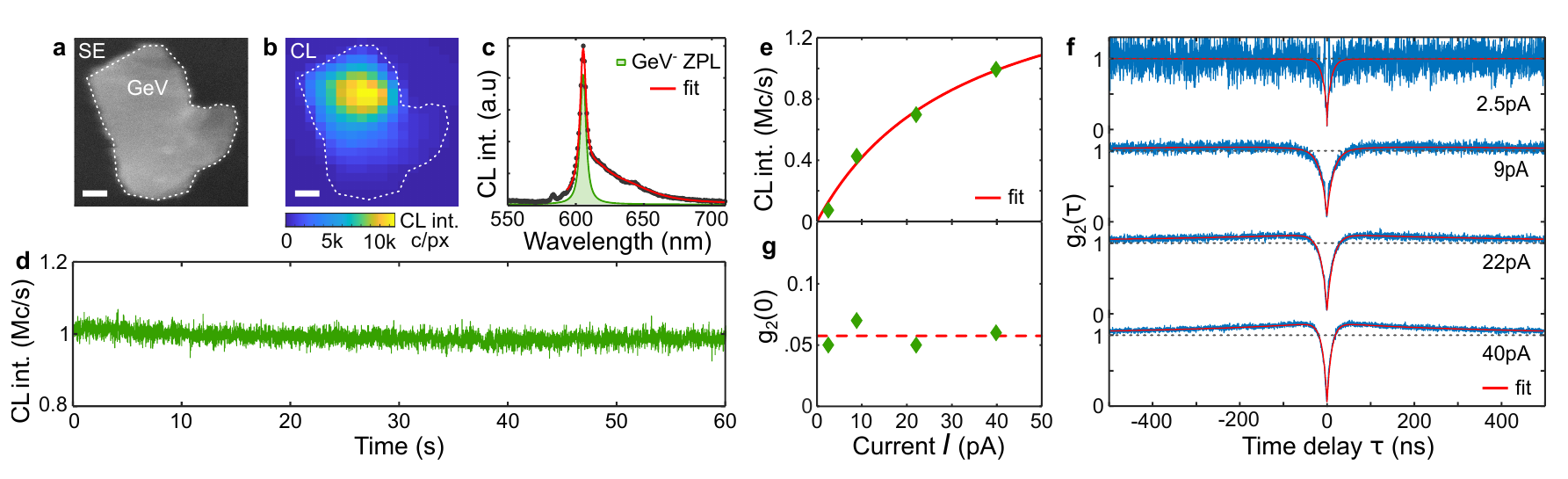}
\caption{
    \textbf{Single-photon emission in CL of an isolated GeV$^-$ center.}
    Secondary electron (SE) image of a diamond crystal (\textbf{a}) and its corresponding CL intensity map (\textbf{b}) revealing the position of a GeV$^-$ center (\SI{40}{pA}, \SI{5}{kV}). The scale bars denote \SI{50}{nm}.
    \textbf{c}~CL spectrum of the localized GeV$^-$ center with ZPL at \SI{605}{nm} (acquired at 40~pA).
    \textbf{d}~CL intensity time trace measured at \SI{40}{pA} presents stable and bright emission at $\sim10^6$ counts per second.
    \textbf{e}~CL intensity saturation of the GeV$^-$ center.
    \textbf{f}~Intensity-correlation histograms of the GeV$^-$ center reveals the single-photon nature of generated emission at all currents.
    \textbf{g}~Single-photon purity $g_2(0)=0.06\pm0.01$ is independent from the electron-beam current.	
	}
\label{fig-single}
\end{figure*}

The increase of electron-beam current prevents the ensemble synchronization, which manifests in the suppression of the zero-correlation peak in the second-order intensity-correlation histogram $g_2(\tau)$. 
Ultimately, $g_2(0)$ converges to 1 at high electron-beam currents for GeV$^-$ and SiV$^-$ ensembles (flat histograms in Fig.~\ref{fig-many}c), as in the case of photo-excitation of the same crystals (see {SI Fig.~S3}), confirming a high number of color centers within the diamond crystals ($N>100$).
Fig.~\ref{fig-many}f summarizes the transition from photon bunching to Poissonian photon statistics in an ensemble with a large number of emitters. 
As the electron-beam current is increased, the incoming electrons arrive closer in time and the photon packets emitted by such an ensemble become {eventually indistinguishable}, ultimately destroying the emission synchronization to the point where the Poissonian distribution is reached, i.e. $g_2(0) = 1$.
This $g_2(0)$-dynamics can be fitted with a function inversely proportional to the electron-beam current~$I$:
\begin{equation}
g_2(\tau=0,I,N) = \left(1-\frac{1}{N}\right)\left(1+\frac{I_0}{I}\right), 
\label{iquation}
\end{equation}
where $I_0$ is the minimal current required for the high-energy interband excitation~\cite{Meuret:2015,Yuge:2022}. 
Therefore, an ensemble of $N$ emitters asymptotically approaches $g_2(0) \rightarrow 1-1/N$ as in the case of photo-excitation.
The horizontal shift of $I$-dependency in Fig.~\ref{fig-many}f indicates a larger $I_0$ required for excitation of SiV$^-$ color centers in comparison with GeV$^-$ ones, which we attribute to their lower quantum yield~\cite{Bradac:2019}.
This $I$-dependent photon bunching in CL has been also observed in various solid-state systems, as for example nitrogen vacancy (NV) centers in diamond, quantum wells in indium gallium nitride, and hexagonal boron nitride (h-BN) encapsulated tungsten disulfide (WS$_2$) monolayers~\cite{Meuret:2015,SolaGarcia2021,FiedlerWS2}.

\subsection{Individual GeV$^-$ center}

In the limit of only one color center, the photon statistics is independent of the electron-beam current, which we demonstrate by the example of an individual GeV$^-$ center in an isolated crystal. 
Fig.~\ref{fig-single}a presents an SEM image of the crystal under investigation, while Fig.~\ref{fig-single}b shows the corresponding CL intensity map ({40~pA, 5~kV}). 
The bright spot in Fig.~\ref{fig-single}b reveals the position of a GeV$^-$ center, which can be localized well below the optical diffraction limit down to $\pm$\SI{5}{nm}.
Fig.~\ref{fig-single}c presents the CL spectrum of the localized bright spot confirming the GeV$^-$ origin of emission. 
The spectral fit highlights the ZPL at \SI{605}{nm} of the GeV$^-$ center with a linewidth of \SI{5}{nm} (\SI{17}{meV}). 
Electron-beam excitation of the localized GeV$^-$ center results in a robust and bright emission producing $\sim10^6$~photon counts per second (not corrected for collection efficiency), which is presented in the CL intensity time trace in Fig.~\ref{fig-single}d.
Measurements of this GeV$^-$ center at increasing electron-beam currents reveal a typical intensity saturation dynamics (Fig.~\ref{fig-single}e), which we fit to obtain a saturation intensity of $1.45\times 10^6$~photon counts per second (see {SI}).

Fig.~\ref{fig-single}f presents intensity-correlation histograms measured at increasing electron-beam currents. 
The increase of electron-beam current to \SI{40}{pA} causes photon bunching at non-zero correlation times, which indicates the presence of a metastable (shelving) state~\cite{Nahra2021}.
We summarize the results in Fig.~\ref{fig-single}g, where we plot $g_2(0)$ extracted from the fit of intensity-correlation histograms from Fig.~\ref{fig-single}f as a function of electron-beam current (see {SI}). 
As expected, the antibunching behavior -- $g_2(0) = 0.06\pm0.01$ -- is observed at all applied electron-beam currents, confirming that this GeV$^-$ center is indeed a single-photon emitter with sub-Poissonian photon statistics. 

\begin{figure*}
\includegraphics[width=0.70\linewidth]{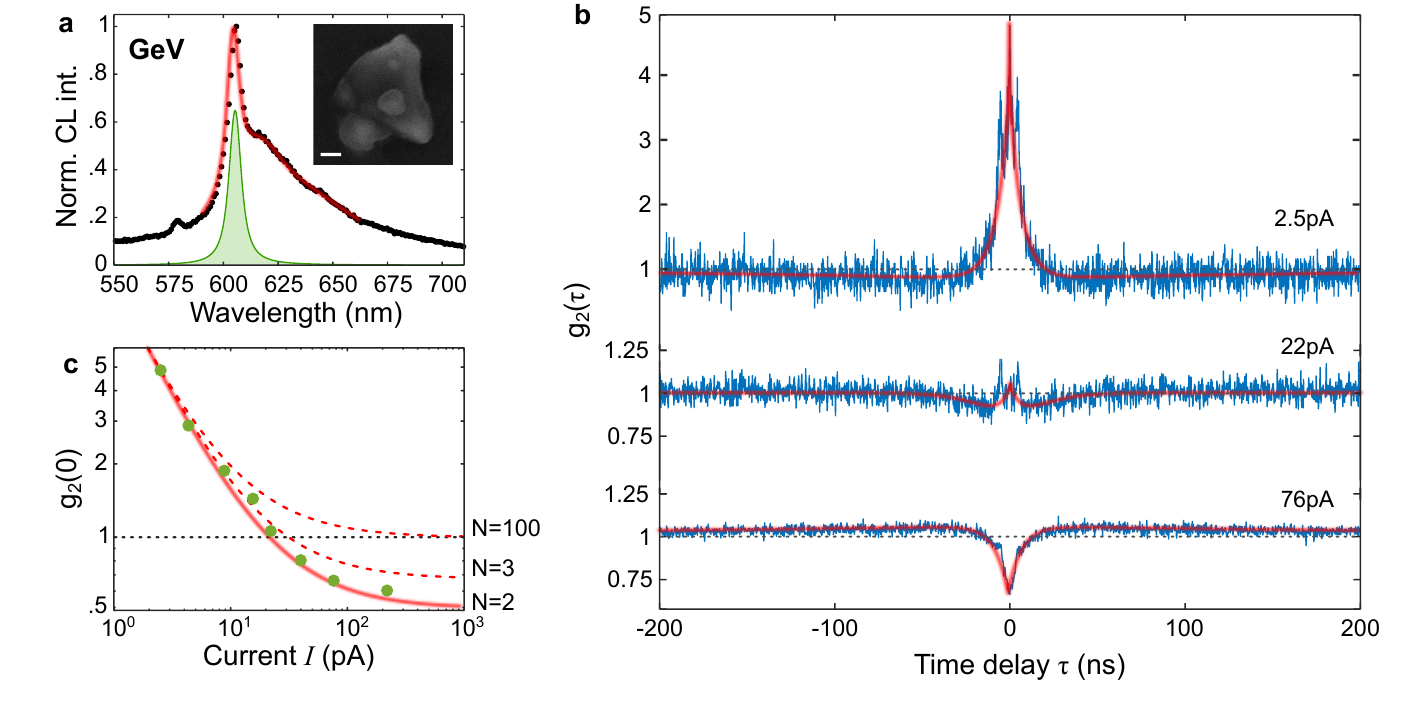}
\caption{\textbf{Cathodoluminescence of an ensemble with two color centers.}
	\textbf{a}~CL spectrum of a nano-diamond with few GeV$^-$ centers emitting at \SI{606}{nm}. The inset shows the SE image of the nano-diamond. The scale bar denotes \SI{100}{nm}.
	\textbf{b}~Second-order intensity-correlation histograms $g_2(\tau)$ measured at increasing electron-beam currents, revealing the transition from super to sub-Poissonian statistics. Note the different vertical scale for the low and high electron-beam currents.  
	\textbf{c}~Extracted $g_2(0)$ values from panel \textbf{b} are plotted as a function of applied electron-beam current $I$. The $I$-dependency is fitted with Eq.~\ref{iquation} (solid red) converging to $g_2(0)=0.5$, indicating that the emitting ensemble consists of $N=2$ color centers. To illustrate the certainty in extracting $N$, the red dashed lines show the expected $I$-dependency in the cases of $N=3$ and $N=100$ emitters.
	}
\label{fig-few}
\end{figure*}

\subsection{Ensemble of two GeV$^-$ centers}

Finally, we turn to discuss the intermediate case of a pair of only two GeV$^-$ centers in the excitation volume of electron beam.
The inset of Fig.~\ref{fig-few}a shows an SEM image of a crystal, which under exposure to the electron beam ({40~pA, 5~kV}) reveals a GeV$^-$ fingerprint in the PL spectrum. 
We further test this crystal at increasing electron-beam currents and record intensity-correlation histograms $g_2(\tau,I)$ (Fig.~\ref{fig-few}b). 
At low electron-beam currents, we observe photon bunching with $g_2(0)=5\pm X$, similar to the case of a large ensemble of GeV$^-$ centers  (see Fig.~\ref{fig-many}c). 
However, when $I$ is increased, the bunching factor does not only decrease as previously shown, but a gradual transition from super to sub-Poissonian behavior is found.

We summarize the changes in photon statistics at zero-correlation time as a function of current in Fig.~\ref{fig-few}c.
Fitting the transition in the $g_2(0,I)$ dynamics with Eq.~\ref{iquation} results in the number of involved GeV$^-$~color centers of $N=2$. 
Indeed, in the case of only two emitters, Eq.~\ref{iquation} converges to $g_2(0)=0.5$ at high electron-beam currents. 
{Additionally, we model the $I$-dependency for ensembles of 3 and 100 color centers (red dashed lines), which at high electron-beam current approach $g_2(0)$ values of 2/3 and 1. }
Such a transition from photon bunching to antibunching behavior of emitters, namely NV color centers in a diamond crystal, has been theoretically explored by Meuret~\emph{et al.}~\cite{Meuret:2015}. To the best of our knowledge, our work experimentally confirms the change from super to sub-Poissonian photon statistics for the first time. The electron-beam excitation allows for control of photon statistics through excitation synchronization of an ensemble of quantum emitters simply with change in current, which can, in-principle, be obtained by excitation with a pulsed laser with pulses randomly separated in time mimicking the electron distribution at various currents~\cite{Meuret:2015}. 

\section{Conclusion}

In conclusion, we report electron-beam excitation of GeV$^-$ and SiV$^-$ color centers of diamond crystals and manipulation of photon-emission statistics. 
By means of varying the electron-beam current, we demonstrated both photon bunching and antibunching behavior of color centers in three regimes, (i) a large ensemble of centers, (ii) a single-photon emitter, and (iii) a few color centers.
While the electron-beam excitation of a single emitter resulted in an antibunching of \num{0.06} regardless of the applied current, a large ensemble of GeV$^-$~centers exhibited photon bunching at low electron-beam currents below and converged to \num{1} at high $I$. 
We showed that the emission of an ensemble with $N>1$ color centers can be synchronized by incoming electrons that are sufficiently separated in time. 
In the intermediate case of only a few GeV$^-$~color centers in an ensemble, via tuning the electron-beam current one can control the synchronization of emitters, which allows to manipulate the photon output from super to sub-Poissonian statistics.  
Ultimately, the experimental confirmation of the tunability of such a source opens up new opportunities for room-temperature applications in quantum computing information technologies.

\section{Supporting Information}
Supporting Information includes details on synthesis of color centers in diamond crystals, cathodoluminescence spectroscopy, optical spectroscopy, scanning-electron microscopy, Monte Carlo simulations, and fitting procedures.
The data that support the findings of this study are available from the corresponding author upon reasonable request.

\section{Author contributions}

S.~M. and S.~F. performed the CL and the HBT measurements. S.~F. performed the Monte Carlo simulations of electron trajectories. E.~A.~E., L.~F.~K., V.~D., and V.~A. fabricated the diamond nano-crystals, while S.~F. prepared the samples. D.~K. and S.~K. optically characterized the samples. S.~M. and S.~F. analyzed the data, while the results were discussed by all authors, and the writing of the manuscript was done in a joint effort. C.~W., S.~I.~B. and N.~A.~M. supervised the project. 

\newpage

\section{Author information}

S. Fiedler\,\orcidlink{0000-0002-7753-0814} \href{https://orcid.org/0000-0002-7753-0814}{orcid.org/0000-0002-7753-0814}

S. Morozov\,\orcidlink{0000-0002-5415-326X} \href{https://orcid.org/0000-0002-5415-326X}{orcid.org/0000-0002-5415-326X}

D. Komisar\,\orcidlink{0000-0001-8856-7586} \href{https://orcid.org/0000-0001-8856-7586}{orcid.org/0000-0001-8856-7586}

E. A. Ekimov\,\orcidlink{0000-0001-7644-0078} \href{https://orcid.org/0000-0001-7644-0078}{orcid.org/0000-0001-7644-0078}

L. F. Kulikova\,\orcidlink{0000-0002-9070-0590} \href{https://orcid.org/0000-0002-9070-0590}{orcid.org/0000-0002-9070-0590}

V.~A.~Davydov\,\orcidlink{0000-0002-8702-0340} \href{https://orcid.org/0000-0002-8702-0340}{orcid.org/0000-0002-8702-0340}

V. N. Agafonov\,\orcidlink{0000-0001-5770-1252} \href{https://orcid.org/0000-0001-5770-1252}{orcid.org/0000-0001-5770-1252}

S.~Kumar\,\orcidlink{0000-0001-5795-0910} \href{https://orcid.org/0000-0001-5795-0910}{orcid.org/0000-0001-5795-0910}

C. Wolff\,\orcidlink{0000-0002-5759-6779} \href{https://orcid.org/0000-0002-5759-6779}{orcid.org/0000-0002-5759-6779}

S. I. Bozhevolnyi\,\orcidlink{0000-0002-0393-4859} \href{https://orcid.org/0000-0002-0393-4859}{orcid.org/0000-0002-0393-4859}

N. A. Mortensen\,\orcidlink{0000-0001-7936-6264} \href{https://orcid.org/0000-0001-7936-6264}{orcid.org/0000-0001-7936-6264}

\newpage
\begin{figure*}
  \centering
  \includegraphics[width=8.25cm]{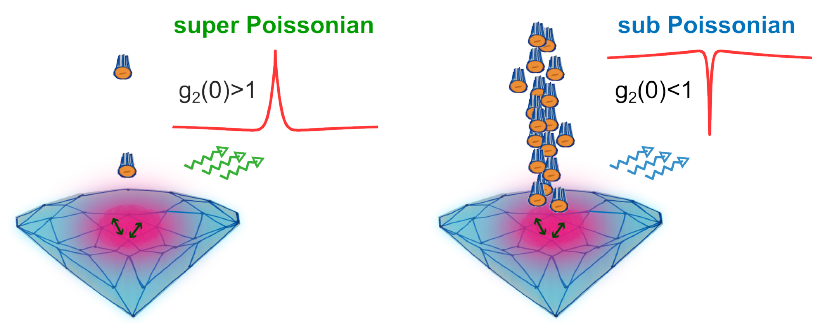}
  \caption*{\textbf{Graphical TOC entry.} An ensemble of 2 color centers in diamond crystal generates sub and super-Poissonian light under electron-beam excitation.}
\end{figure*}

\end{document}